\documentclass[10pt]{article}
\usepackage{amsfonts}
\usepackage{epsfig,cite}
\begin{document}
\begin{center}
%%%%%%%%%%%%%%%%%%%%%%%%%%%%%%%%%%%%%%%%%%%%%%%%%%%%%%%%%%%%%%%%%%%%%%%%%%%
{\Large\bf Resonances contribution to two-photon exchange effects and possible large forward-backward asymmetry in $e^+  e^- \leftrightarrow p  \bar{p}$ }\\
%%%%%%%%%%%%%%%%%%%%%%%%%%%%%%%%%%%%%%%%%%%%%%%%%%%%%%%%%%%%%%%%%%%%%%%%%%%
\vspace*{1cm}
%%%%%%%%%%%%%%%%%%%%%%%%%%%%%%%%%%%%%%%%%%%%%%%%%%%%%%%%%%%%%%%%%%%%%%%%%%%
Hai-Qing Zhou $^{1,3} \footnote{E-mail: zhouhq@mail.ihep.ac.cn}$, Bing-Song Zou $^{2,3}$\\
\vspace{0.3cm} %
{\ $^1$ Department of Physics, Southeast University,
 Nanjing\ 211189,\ P. R. China}\\
{\ $^2$ Institute of High Energy Physics, CAS, Beijing\ 100049, \ P. R. China}\\
{\ $^3$ Theoretical Physics Center for Science Facilities, CAS, Beijing\ 100049, \ P. R. China}\\
%%%%%%%%%%%%%%%%%%%%%%%%%%%%%%%%%%%%%%%%%%%%%%%%%%%%%%%%%%%%%%%%%%%%%%%%%%%
\vspace*{1cm}
\end{center}
%%%%%%%%%%%%%%%%%%%%%%%%%%%%%%%%%%%%%%%%%%%%%%%%%%%%%%%%%%%%%%%%%%%%%%%%%%%
\begin{abstract}
The resonances ($\eta_c,\chi_{c0,c2}$) contribution to two-photon exchange (TPE) effects in $e^+  e^- \leftrightarrow p  \bar{p}$ is calculated in a simple hadronic model.
The calculation shows the TPE contributions by resonance $\chi_{c2}$, which are dependent on the unknown phases $\phi_{E,M}$ of proton's time-like form factors $G_{E,M}$, are much larger than the TPE contributions by non-resonance and are comparable with measurement precision of coming PANDA detector at $\sqrt{s} \sim M_{\chi_{c2}}$ for most $\phi_{E,M}$.
\end{abstract}
\section{Introduction}
The structure of nucleon is one of the most important topics in the hadronic physics. The electromagnetic interaction provides a clean method to measure such a structure. One of the most important measurement is to determine the electromagnetic form factors $G_E(Q^2)$ and $G_M(Q^2)$ of nucleon. The measurements of $R=\mu_p G_E(Q^2)/G_M(Q^2)$ at space-like region by polarized method and Roesenbluth method \cite{Ex-FFs} indicate it is not trivial to extract the form factors directly from the angle dependence of cross section. Since then, the two-photon exchange (TPE) effects attract many interest. In the literature, the TPE contributions in $ep \rightarrow ep $ and $e^+  e^-  \leftrightarrow p  \bar{p}$ have been estimated by some model dependent methods \cite{TPE-calculation-hadronic,TPE-calculation-GPD,TPE-calculation-dispersion,TPE-calculation-pQCD,TPA-hadronic-model,TPA-pQCD} and model independent analysis \cite{TPE-analysis-1,TPE-analysis-2,TPE-analysis-3,TPE-analysis-4,TPE-analysis-5,TPA-analysis} (see recent review articles \cite{TPE-review}). Experimentally, to detect the TPE effects directly, the measurements of $R_{e^+e^-}$ defined as the ratio of $e^{+}p$ to $e^{-}p$ differential cross sections are proposed by VEPP-3\cite{Ex-R-VEPP}, JLab\cite{Ex-R-JLab} and OLYMPUS\cite{Ex-R-OLYMPUS}, and the measurement of forward-backward asymmetry in $p  \bar{p} \rightarrow e^+  e^-  $ is proposed by PANDA\cite{Ex-FF-PANDA}. In this work, we estimate the TPE contributions in the unpolarized processes $e^+  e^- \leftrightarrow p \bar{p}$ when $\sqrt{s}$ lies in the resonances $\eta_c,\chi_{c0,c2}$ region. We arrange our discussion as follows: in Section 2, we review the contributions of vector resonances ($\psi,\psi'$) in $e^+  e^- \rightarrow p \bar{p}$ by one-photon exchange (OPE); in Section 3, we discuss the TPE contributions by $\eta_c,\chi_{c0,c2}$  and in Section 4 we present the numerical results and give a discussion.
\section{Resonances contribution in $e^+  e^- \rightarrow p \bar{p}$ by OPE }
\begin{figure}[t]
\centerline{\epsfxsize 3.0 truein\epsfbox{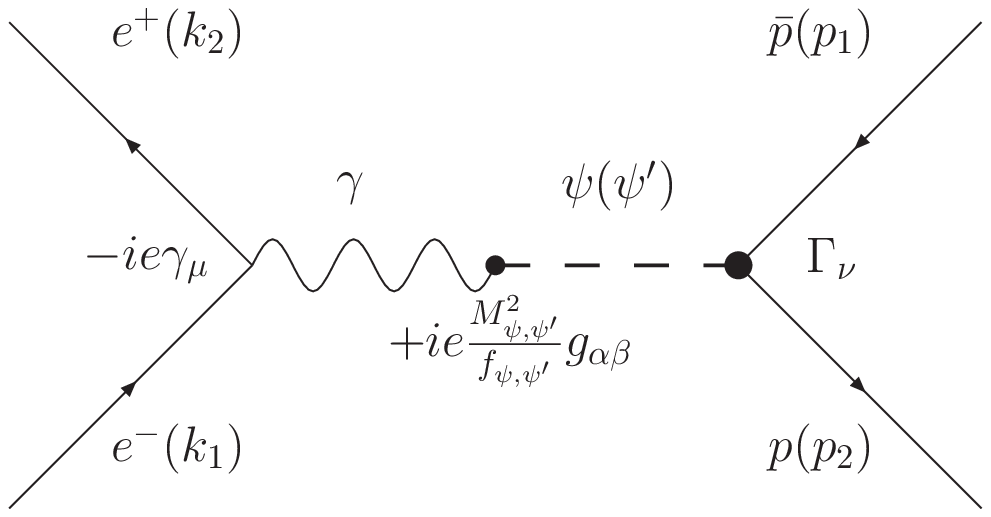}\epsfxsize 3.0 truein\epsfbox{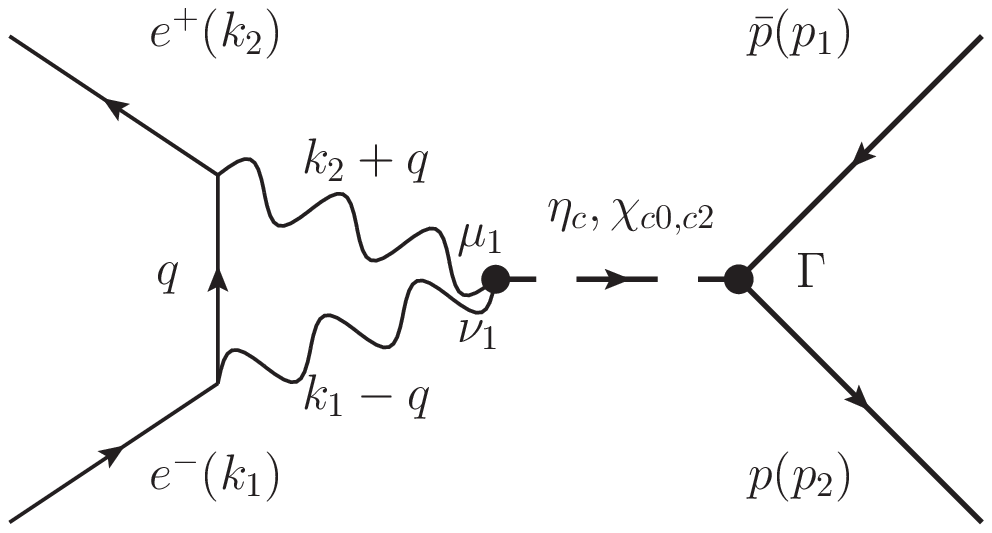}}
\caption{Left panel: diagrams for $1^{--}$ resonances ($\psi$,$\psi'$) contributions in $e^+  e^-
\rightarrow p \bar{p}$ by one photon annihilation; right panel: diagrams for $0^{-+},0^{++},2^{++}$ resonances ($\eta_c,\chi_{c0,c2}$) contributions in $e^+ e^- \rightarrow p \bar{p}$ by two photon annihilating.}
\label{OPE-TPE}
\end{figure}
Considering the process $e^+(k_2) + e^-(k_1) \rightarrow p(p_2) +
\bar{p}(p_1)$ at tree level, by the
vector-meson-dominance model, the $1^{--}$ resonances contributions can be
described by diagrams in the left panel of Fig.\ref{OPE-TPE}, where we limit our discussion in the charmonium region. The couplings of $1^{--}$ charmonium
$\psi,\psi'$ with $p\bar{p}$ are described as in Ref.\cite{T.Barnes2008-vector-particle-coupling} with a little different notation
\begin{eqnarray}
\Gamma_{\nu}^{(\psi
p\bar{p},\psi'p\bar{p})}=-i(g_{\psi,\psi'}\gamma_{\nu}+\frac{i\kappa_{\psi,\psi'}}{2M_N}\sigma_{\nu\rho}q^{\rho}),
\end{eqnarray}
where $M_N=0.938$ GeV is the mass of proton,
$q=(\sqrt{s},0,0,0)$ in the c.m. frame \cite{T.Barnes2008-vector-particle-coupling-corrected}. By these Born diagrams, the unpolarized differential cross section
in the c.m. frame can be written as
\begin{eqnarray}
\frac{d\sigma^0}{d\Omega}&=&\frac{\alpha_{em}^2M_\psi^4(1-4M_N^2/s)^{1/2}}{4s^2}\frac{[4M_N^2|G_E^{\psi}|^2(1-\cos^2\theta)
+s|G_M^{\psi}|^2(1+\cos^2\theta)]}{f_\psi^2((s-M_\psi^2)^2+\Gamma_{\psi}^2M_\psi^2)}+\nonumber\\
&&\frac{\alpha_{em}^2M_{\psi'}^4(1-4M_N^2/s)^{1/2}}{4s^2}\frac{[4M_N^2|G_E^{\psi'}|^2(1-\cos^2\theta)
+s|G_M^{\psi'}|^2(1+\cos^2\theta)]}{f_{\psi'}^2((s-M_{\psi'}^2)^2+\Gamma_{\psi'}^2M_{\psi'}^2)}+\nonumber\\
&&\frac{\alpha_{em}^2M_\psi^2M_{\psi'}^2(1-4M_N^2/s)^{1/2}}{4s^2}2Re\{\frac{4M_N^2G_E^{\psi}G_E^{\psi'*}(1-\cos^2\theta)+
sG_M^{\psi}G_M^{\psi'*}(1+\cos^2\theta)}{f_\psi
f_{\psi'}^{*}(s-M_\psi^2+i\Gamma_\psi
M_\psi)(s-M_{\psi'}^2-i\Gamma_{\psi'} M_{\psi'})}\},
\end{eqnarray}
with $M_{\psi}=3.096$ GeV, $M_\psi'=3.686$ GeV the masses of $\psi,\psi'$, $\theta$ the scattering angle in the c.m. frame, and $G_E^{\psi,\psi'}=g_{\psi,\psi'}+\kappa_{\psi,\psi'} s/4M_N^2, ~~
G_M^{\psi,\psi'}=g_{\psi,\psi'}+\kappa_{\psi,\psi'}$. Since we limit the
discussion in the charmonium region, it is a good
approximation to treat the couplings $f_{\psi,\psi'}, G_{E,M}^{\psi,\psi'}$ as constants which are constrained by \cite{T.Barnes2008-vector-particle-coupling}
\begin{eqnarray}
&&\Gamma_{\psi,\psi'\rightarrow
p\bar{p}}=\frac{(1-4M_N^2/M_{\psi,psi'}^2)^{1/2}}{12 \pi
M_{\psi,\psi'}}(2M_N^2|G_E^{\psi,\psi'}|^2+M_{\psi,\psi'}^2|G_M^{\psi,\psi'}|^2),\nonumber\\
&& \Gamma_{\psi,\psi'\rightarrow
e^{+}e^{-}}=\frac{4\pi\alpha_{em}^2M_{\psi,\psi'}}{3f_{\psi,\psi'}^2},~~~~~\alpha_{\psi,\psi'} = \frac{1-4M_N^2/s|G_E^{\psi,\psi'}/G_M^{\psi,\psi'}|^2}{1+4M_N^2/s|G_E^{\psi,\psi'}/G_M^{\psi,\psi'}|^2},
\end{eqnarray}
with $\alpha_{\psi,\psi'}\approx0.67$ at corresponding $\sqrt{s}=M_{\psi,\psi'}$. In the following calculation, for simplicity, we naively assume that the phases of $G_E^{\psi}$ and $G_E^{\psi'}$, $G_M^{\psi}$ and $G_M^{\psi'}$ are the same, respectively. By these assumptions and the constraint conditions, only two physical parameters $\phi_{M,E}$ which are defined as $G_{E,M}^{\psi,\psi'}=e^{i\phi_{E,M}}|G_{E,M}^{\psi,\psi'}|$ are left as unknown.

\section{Resonances contribution in $e^+  e^- \rightarrow p \bar{p}$ by TPE }
When considering the TPE contributions from other resonances in charmonium region,
the diagrams shown in the right panel of Fig.\ref{OPE-TPE} should be considered, where only $0^{-+},0^{++},2^{++}$ resonances are included since their couplings to $p\bar{p}$ and $2\gamma$ are relatively large. The couplings of $0^{-+},0^{++},2^{++}$ charmonium with two-photon can be described as in \cite{charmonium},
\begin{eqnarray}
\Gamma^{\mu_1\nu_1}_{\gamma\gamma\eta_c}&=&-ig_{\eta_c\gamma\gamma}D\varepsilon^{\mu_1\mu_2\nu_1\nu_2}(k_2+q)_{\mu_2}(k_1-q)_{\nu_2},\nonumber\\
\Gamma^{\mu_1\nu_1}_{\gamma\gamma\chi_{c0}}&=&-ig_{\chi_{c0}\gamma\gamma}D^2\{[(k_1-q)\cdot(k_2+q)g^{\mu_1\nu_1}-(k_1-q)^{\mu_1}(k_2
+q)^{\nu_1}][M_{\chi_{c0}}^2+(k_1-q)\cdot(k_2+q)]\nonumber\\
&&~~~~~~~~~~+(k_2+q)^{\mu_1}(k_2+q)^{\nu_1}(k_1-q)^2+(k_1-q)^{\mu_1}(k_1-q)^{\nu_1}(k_2+q)^2\nonumber\\
&&~~~~~~~~~~-(k_1-q)^2(k_2+q)^2g^{\mu_1\nu_1}-(k_2+q)^{\nu_1}(k_1-q)^{\mu_1}(k_1-q)\cdot(k_2+q)\},\nonumber\\
\Gamma^{\mu_1\nu_1;\mu_2\nu_2}_{\gamma\gamma\chi_{c2}}&=&-ig_{\chi_{c2}\gamma\gamma}D^2\{g^{\mu_1\mu_2}g^{\nu_1\nu_2}(k_1-q)\cdot(k_2+q)
-g^{\mu_1\nu_1}(k_1-q)^{\mu_2}(k_2+q)^{\nu_2}\nonumber\\
&&~~~~~~~~~~-(k_2+q)^{\nu_1}(k_1-q)^{\mu_2}g^{\mu_1\nu_2}
-(k_1-q)^{\mu_1}(k_2+q)^{\mu_2}g^{\nu_1\nu_2} \},
\label{v1}
\end{eqnarray}
where $k_1-q, k_2+q$ are the momentums of photons, $\mu_1, \nu_1$ are the corresponding Lorentz indexes and $D=[(q+k_2/2-k_1/2)^2-m_c^2+i\epsilon]^{-1}$ with $m_c$ the mass of $c$ quark, and we take $m_c=1.5$ GeV in the following calculation.

For the couplings of $0^{-+}, 0^{++}, 2^{++}$ charmonium with $p\overline{p}$, for simplicity we take them the same as those with $e^{+}e^{-}$, which can be described as in \cite{charmonium},
\begin{eqnarray}
\Gamma_{\eta_c p\bar{p}}&=&g_{\eta_cp\bar{p}}\gamma_5,\nonumber\\
\Gamma_{\chi_{c0} p\bar{p}}&=&g_{\chi_{c0}p\bar{p}},\nonumber\\
\Gamma^{\mu\nu}_{\chi_{c2} p\bar{p}}&=&g_{\chi_{c2}p\bar{p}}(p_1-p_2)^{\nu}\gamma^{\mu}.
\label{v2}
\end{eqnarray}
These couplings are constrained by the physical decay width
$\Gamma(\eta_c,\chi_{c0,c2}\rightarrow {\gamma\gamma,p\bar{p}})$ with
\begin{eqnarray}
\Gamma_{\chi_{c2}\rightarrow \gamma\gamma } &=& \frac{g^2_{\gamma\gamma\chi_{c2}}M_{\chi_{c2}}^3}{15(4m_c^2+M_{\chi_2}^2)^2\pi},\nonumber\\
\Gamma_{\chi_{c2}\rightarrow p\bar{p}}&=&\frac{g_{\chi_{c2}p\bar{p}}^2\sqrt{M_{\chi_{c2}}^2-4M_N^2}(3M_{\chi_{c2}}^4-4M_{\chi_{c2}}^2M_N^2-32M_N^4)}{120\pi M_{\chi_{c2}}^2},
\end{eqnarray}
where $M_{\chi_2}=3.556$ GeV is the mass of $\chi_{c2}$, $\Gamma_{\chi_{c2}\rightarrow \gamma\gamma }=1.97$ MeV $\times(2.56\times10^{-4})=5.04\times10^{-4}$ MeV and $\Gamma_{\chi_{c2}\rightarrow p\bar{p}}=1.97$ MeV $\times(7.2\times10^{-5})=1.42\times10^{-4}$ MeV\cite{PDG2010}. This results in
\begin{eqnarray}
|g_{\gamma\gamma\chi_{c2}}|=1.57\times10^{-2},|g_{\chi_{c2}p\bar{p}}|=7.39\times10^{-4}.
\end{eqnarray}
Here, only $g_{\chi_{c2}p\bar{p}}$ and $g_{\gamma\gamma\chi_{c2}}$ are given
since the calculation shows the contributions from
$\eta_c$ and $\chi_{c0}$ are identically zero because of the Dirac structure of corresponding unpolarized differential cross sections. For simplicity, we assume the phases of these two couplings are zero.

The propagator of $\chi_{c2}$ is described as the standard
Breit-Wigner form \cite{propagator-spin-2}
\begin{eqnarray}
S^{\mu_2\nu_2;\rho\omega}_{\chi_{c2}}&=&\frac{-i}{P^2-M_{\chi_{c2}}^2+iM_{\chi_{c2}}{\Gamma_{\chi_{c2}}}}
\{\frac{1}{2}(g^{\mu_2\rho}\frac{P^{\nu_2P^\omega}}{M_{\chi_2}^2}+g^{\nu_2\omega}\frac{P^{\mu_2}P^{\rho}}{M_{\chi_{c2}}^2}
+g^{\mu_2\omega}\frac{P^{\nu_2}P^{\rho}}{M_{\chi_{c2}}^2}+g^{\nu_2\rho}\frac{P^{\mu_2}P^{\omega}}{M_{\chi_{c2}}^2})\nonumber\\
&&+\frac{1}{2}(g^{\mu_2\rho}g^{\nu_2\omega}+g^{\mu_2\omega}g^{\nu_2\rho}-g^{\mu_2\nu_2}g^{\rho\omega})
+\frac{2}{3}(\frac{1}{2}g^{\mu_2\nu_2}-\frac{P^{\mu_2}P^{\nu_2}}{M_{\chi_{c2}}^2})
(\frac{1}{2}g^{\rho\omega}-\frac{P^{\rho}P^{\omega}}{M_{\chi_{c2}}^2})\}.
\label{propagator}
\end{eqnarray}

With Eqs. (\ref{v1},\ref{v2},\ref{propagator}), the corresponding amplitudes for the
diagrams in the right panel of  Fig.\ref{OPE-TPE} can be written down. Their interferences with Born diagrams can be calculated directly and we use the
FeynCalc \cite{FeynCalc} to do the analysis calculation and LoopTools \cite{LoopTools} for the
numerical calculation.

\section{Results}
Since only $\chi_{c2}$ gives the contributions, we limit the discussion in $\sqrt{s}\sim M_{\chi_{c2}}$. To show the TPE contributions, we define
\begin{eqnarray}
\delta &\equiv& \frac{d\sigma^{(1)}/d\Omega}{d\sigma^{(0)}/d\Omega},\nonumber\\
A^{[\theta_0-\theta_1]}&\equiv&\frac{2\pi\int_{\theta_0}^{\theta_1} (d\sigma^{(0)}+d\sigma^{(1)})/d\Omega sin\theta d\theta-2\pi\int_{\pi-\theta_1}^{\pi-\theta_0} (d\sigma^{(0)}+d\sigma^{(1)})/d\Omega sin\theta d\theta}{2\pi\int_{\theta_0}^{\theta_1}(d\sigma^{(0)}+d\sigma^{(1)})/d\Omega sin\theta d\theta+2\pi\int_{\pi-\theta_1}^{\pi-\theta_0}(d\sigma^{(0)}+d\sigma^{(1)}) /d\Omega sin\theta d\theta}\nonumber\\
&=&\frac{\int_{\theta_0}^{\theta_1} d\sigma^{(1)}/d\Omega sin\theta d\theta}{\int_{\theta_0}^{\theta_1}d\sigma^{(0)}/d\Omega sin\theta d\theta},
\end{eqnarray}
with $d\sigma^{(1)}/d\Omega$ the unpolarized differential cross section from the interference of TPE diagrams and Born diagrams. By the crossing symmetry and Lorentz invariance, we have the relation for the amplitudes
$M(e^+(k_2)e^-(k_1) \rightarrow p(p_2)\bar{p}(p_1))=M(p(-p_1)\bar{p}(-p_2) \rightarrow e^+(-k_1)e^-(-k_2))=M(p(p_1)\bar{p}(p_2) \rightarrow e^+(k_1)e^-(k_2))$, which means the ratios of cross sections $\delta, A^{[\theta_0-\theta_1]}$ in $e^+  e^- \rightarrow p  \bar{p}$ are the same as those in $p\bar{p}\rightarrow e^+e^-$.

\begin{figure}[t]
\centerline{\epsfxsize 7 truein\epsfbox{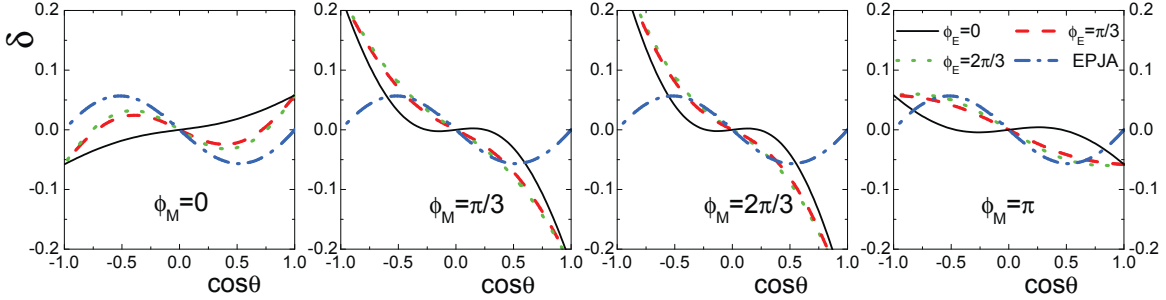}}
\caption{The cos$\theta$ dependence of $\delta$ at $\sqrt{s}=M_{\chi_{c2}}$ by $\chi_{c2}$ contributions to TPE effects. The black solid, red dashed and green dotted curves are corresponding to $\phi_E=0,\pi/3$ and $2/3\pi$, the  blue dashed-dotted curves named as EPJA are explained in the text.}
\label{delta-theta}

\end{figure}
\begin{figure}[t]
\centerline{\epsfxsize 7 truein\epsfbox{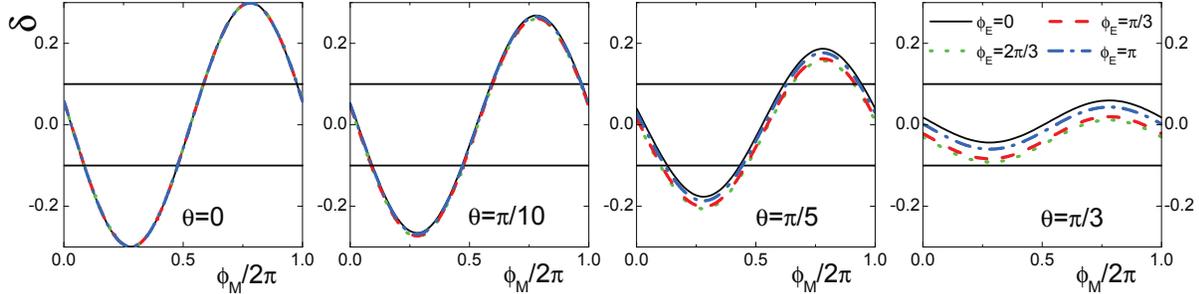}}
\caption{The $\phi_M$ dependence of $\delta$ at $\sqrt{s}=M_{\chi_{c2}}$ by $\chi_{c2}$ contributions to TPE effects. The black solid, red dashed, green dotted and blue dashed-dotted curves are corresponding to $\phi_E=0,\pi/3,2/3\pi$ and $\pi$.}
\label{delta-phi_M}
\end{figure}

The $\theta$ and $\phi_M$ dependence of $\delta$ at $\sqrt{s}=M_{\chi_{c2}}$ are presented in Fig.\ref{delta-theta} and Fig.\ref{delta-phi_M}. The results show the TPE contributions to $\delta$ by $\chi_{c2}$ are almost independent on $\phi_E$, especially at small $\theta$, while are strongly dependent on $\theta$ and $\phi_M$. The corrections $\delta$ are odd functions of $\cos\theta$, which is a general property of $2\gamma$ annihilation effects. The most interesting property of Fig.\ref{delta-phi_M} is that the absolute corrections $|\delta|$ are larger than 10\% at small $\theta$ ($\theta \leq \pi/10$) for almost all $\phi_M$.

To compare with the measurement precision of coming PANDA detector, we present the corresponding $\delta$ used by the PANDA simulations \cite{PANDA-TPE}  as the curves labeled EPJA in Fig.\ref{delta-theta}. By the definition of our $\delta$ and TPE contribution Eq.(14) of Ref.\cite{PANDA-TPE}, we have
\begin{eqnarray}
\delta=\frac{2\sqrt{\tau(\tau+1)}(G_E/\tau-G_M)F_3cos(\theta)sin^2(\theta)}{G_M^2(1+cos^2(\theta))+(G_E^2/\tau) sin^2(\theta)},
\end{eqnarray}
with $\tau=s/4M_N^2$ and $F_3$ defined by Eq.(14) of Ref.\cite{PANDA-TPE}. The curve of $\delta$ named as EPJA in Fig.\ref{delta-theta} is corresponding to the case $F_3/G_M=5\%$ assuming $G_E=G_M$. The results in Fig.\ref{delta-theta} show that the realistic $\delta$ in the small $\theta$ ($|cos\theta|>0.8$) are much larger than $\delta$(EPJA) for most $\phi_M$. In the region where PANDA detector works, the realistic $\delta$ are larger than $\delta$(EPJA) in the region  $0.5<|cos\theta|<0.8$ and  comparable with $\delta$(EPJA) in the region $|cos\theta|<0.5$ for some $\phi_M$, for example $\phi_M=1/3\pi,2/3\pi$.

The full results show that the TPE contributions to $\delta$ from $\chi_{c2}$ are much larger than the usual TPE effects from non-resonance contributions. The latter are usually less than 1\% at small $\theta$ \cite{TPA-hadronic-model,TPA-pQCD} and is a challenge to be observed by PANDA \cite{TPA-pQCD}, while the former are comparable with the measurement precision of PANDA at $\sqrt{s}\sim M_{\chi_{c2}}$. This suggests that the coming PANDA experiment may detect the direct TPE effects at $\sqrt{s}\sim M_{\chi_{c2}}\pm 0.01$ GeV. And outside of this region, the TPE contributions from $\chi_2$ are as small as 1\% at the small $\theta$ ($\theta<\pi/10$) and the corrections to the measurements of $G_{M,E}$ will be very small \cite{PANDA-TPE}.

\begin{figure}[t]
\centerline{\epsfxsize 3.0 truein\epsfbox{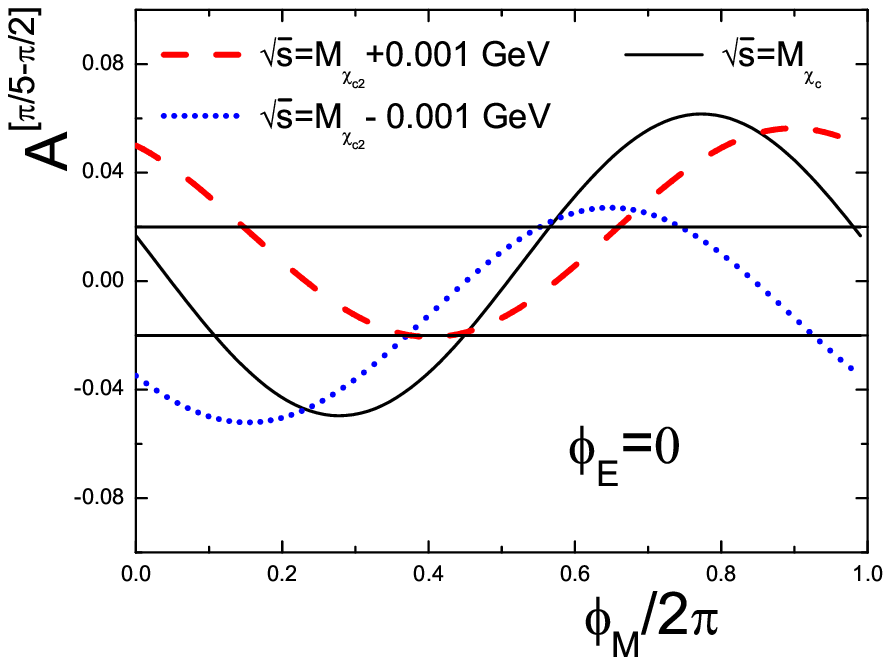}\epsfxsize 3.0 truein\epsfbox{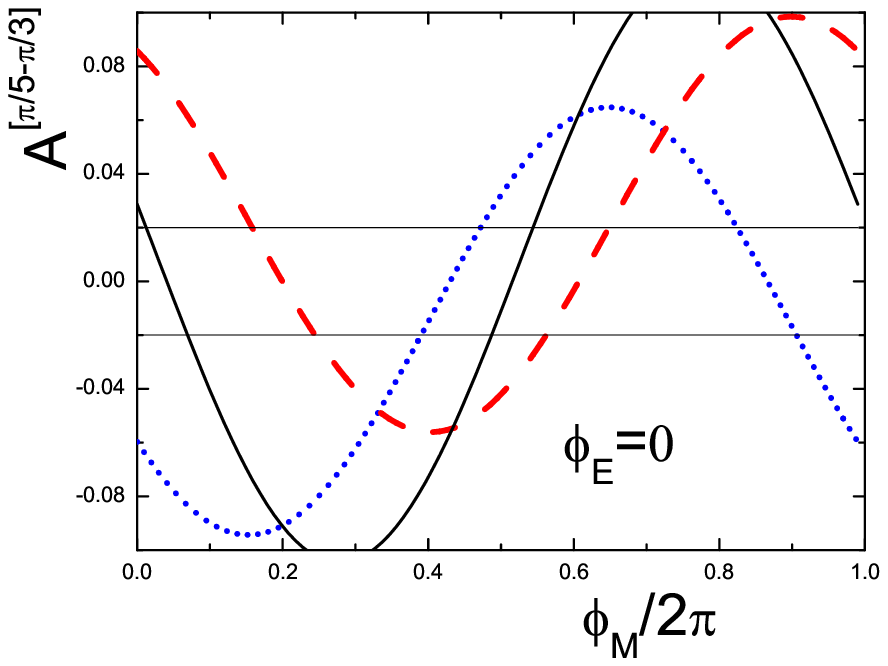}}
\caption{The $\phi_M$ dependence of forward-backward asymmetry at $\sqrt{s}\sim M_{\chi_{c2}}$ with $\phi_E=0$. The red dashed, black solid and blue dotted curves are corresponding to $\sqrt{s}=M_{\chi_{c2}}+0.001$ GeV, $M_{\chi_{c2}}$ and $M_{\chi_{c2}}-0.001$ GeV.}
\label{A-phi}
\end{figure}

\begin{figure}[t]
\centerline{\epsfxsize 3.5 truein\epsfbox{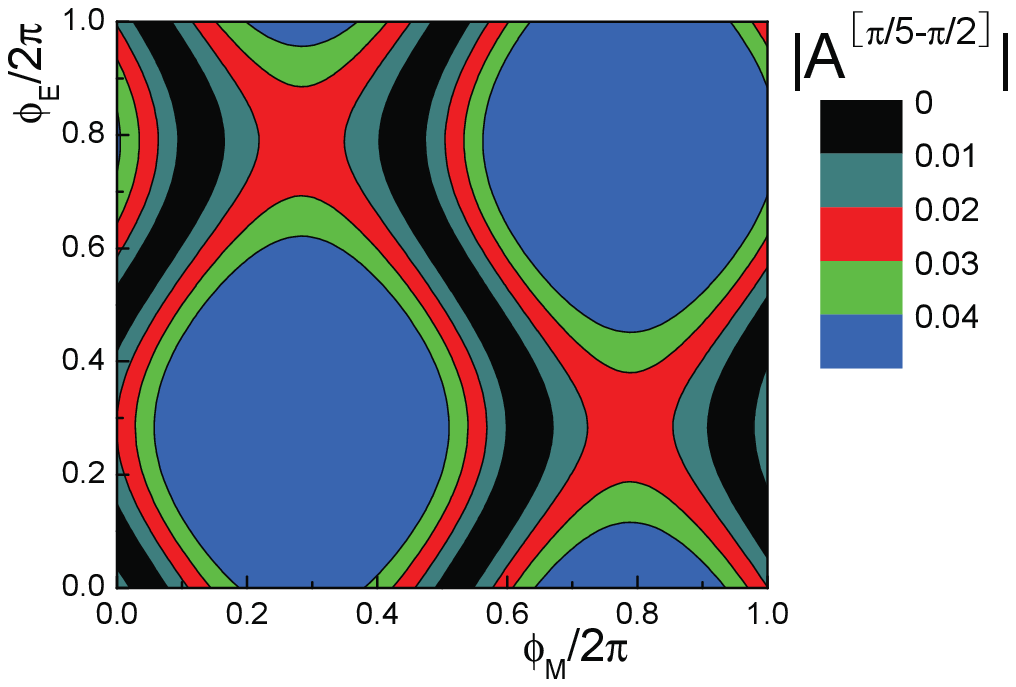}\epsfxsize 3.5 truein\epsfbox{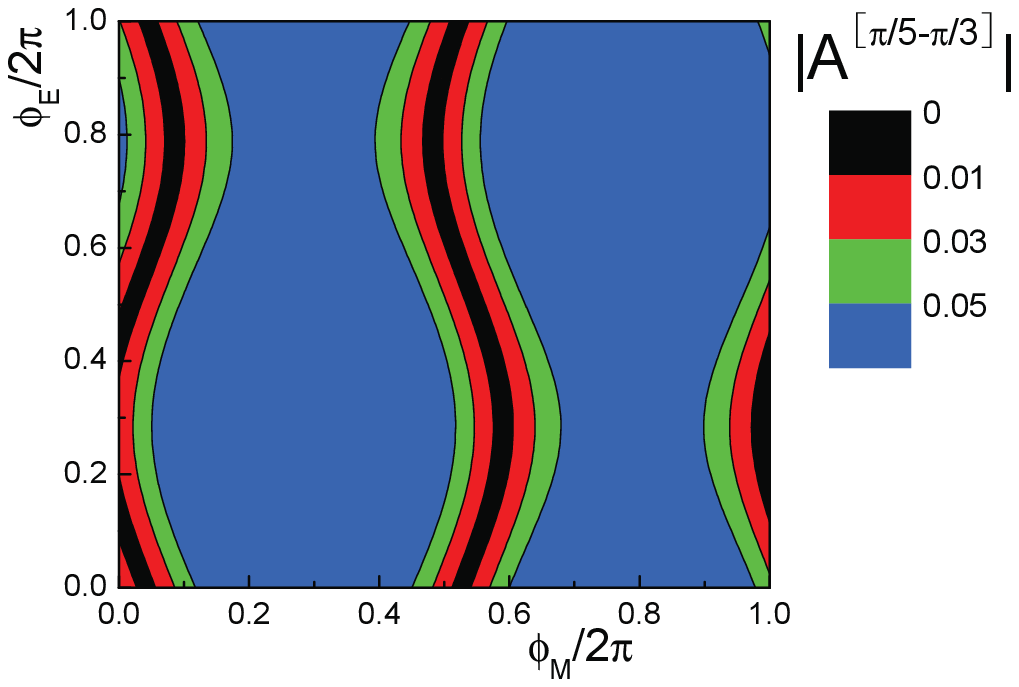}}
\caption{The $\phi_{M,E}$ dependence of absolute forward-backward asymmetry at $\sqrt{s}= M_{\chi_{c2}}$ in contour form.}
\label{A-con}
\end{figure}

Fig.\ref{A-phi} displays the curves of forward-backward asymmetry $A^{[\theta_0-\theta_1]}$ vs. $\phi_M$ at $\sqrt{s}=M_{\chi_{c2}},M_{\chi_{c2}}\pm0.001$ GeV with $\phi_E=0$ as an example. The left panel of Fig.\ref{A-phi} shows that the forward-backward asymmetry with $\theta_0=\pi/5,\theta_1=\pi/2$ is larger than 2\% for almost all $\phi_M$ when combining the three $\sqrt{s}$ points. The right panel of Fig.\ref{A-phi} shows that the forward-backward asymmetry with smaller $\theta_1=\pi/3$ is much enhanced. Fig.\ref{A-con} displays the $\phi_{E,M}$ dependence of $A^{[\pi/5-\pi/3]}$ and $A^{[\pi/5-\pi/2]}$ in contour form, which shows the properties of forward-backward asymmetry in the parameter space more clearly. By the simulation of PANDA detector \cite{PANDA-TPE}, the reconstructed number of counts at $s=12.9$ GeV$^2$ ($\sim M_{\chi_{c2}}^2$) is about $10^3$ for $|cos\theta|<0.8$ (corresponding to $\theta>1/5\pi$). The left panel of Fig.\ref{A-con} shows that the number of asymmetry events for $\theta \subseteq [\pi/5-\pi/2]$ at $\sqrt{s}\sim M_{\chi_{c2}}$ is larger than 20 for almost all $\phi_{E,M}$ and larger than 40 most $\phi_{E,M}$.

\begin{figure}[t]
\centerline{\epsfxsize 7.5 truein\epsfbox{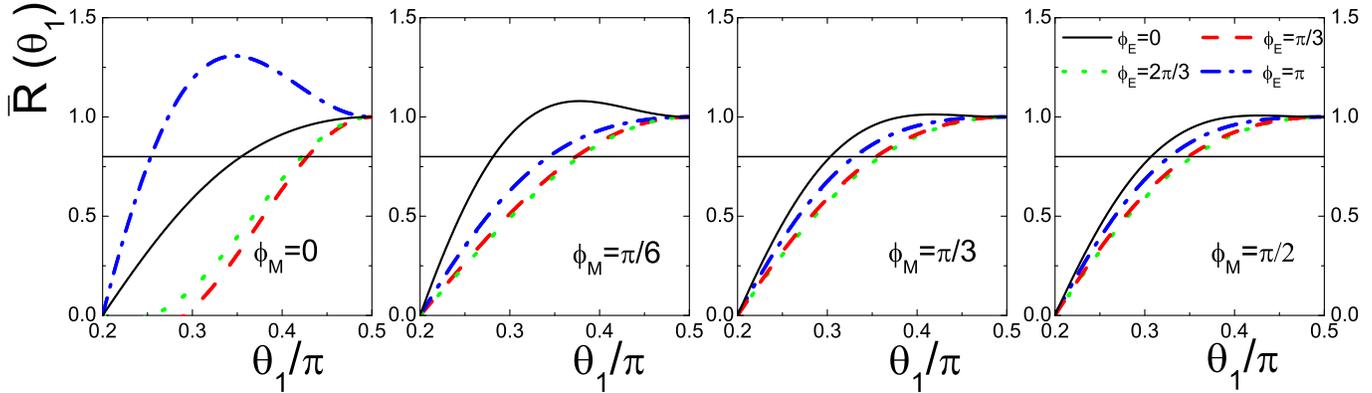}}
\caption{The $\theta_1$ dependence of $R$ at $\sqrt{s}= M_{\chi_{c2}}$ with $\theta_0=\pi/5$.  The black solid, red dashed, green dotted and blue dashed-dotted curves are corresponding to $\phi_E=0,\pi/3,2/3\pi$ and $\pi$.}
\label{RA-theta}
\end{figure}

To show the property of asymmetry events number, we define the ratio $\bar{R}$ as
\begin{eqnarray}
\bar{R}\equiv\frac{\int_{\theta_0}^{\theta_1} d\sigma^{(1)}/d\Omega sin\theta d\theta}{\int_{\theta_0}^{\pi/2}d\sigma^{(1)}/d\Omega sin\theta d\theta}.
\end{eqnarray}
The angle $\theta_1$ dependence of ratio $\bar{R}$ at $\sqrt{s}=M_{\chi_{c2}}$ with $\theta_0=\pi/5$ is presented in  Fig.\ref{RA-theta} which shows two interesting properties: (1) $\bar{R}$ is almost independent on $\phi_{E,M}$ except for at very small $\phi_M$. The special behavior of $\bar{R}$ at very small $\phi_M$ is corresponding to the property of $\delta$ at $\phi_M=0$ in Fig.\ref{delta-theta}, which is not monotonic function of $cos\theta$ in the region $\theta \subseteq [\pi/5,\pi/2]$; (2) $\bar{R}$ reaches about 80\% for $\theta_1=\pi/3$ and 90\% for $\theta_1=2\pi/5$. These two properties indicate that whatever $\phi_{E,M}$ are, for experiments, most of the forward-backward asymmetry events lie in $\theta\subseteq[\pi/5,\pi/3]/[2\pi/3,4\pi/5]$ when the detector works in the region $|cos\theta|<0.8$.

In conclusion, our calculation shows the resonance $\chi_{c2}$ contributions to TPE effects may give large corrections  in $e^+e^- \leftrightarrow p\overline{p}$ when $\sqrt{s} \sim M_{\chi_{c2}}$. The results suggest that the coming experiment PANDA may detect the direct TPE effects in small scattering angle region such as $\theta\subseteq[\pi/5,\pi/3]$ at $\sqrt{s}\sim M_{\chi_{c2}}$. For lower energies around 2.3 GeV, the contributions from resonances of light flavor quarks should also play similar important role  in $e^+e^- \leftrightarrow p\bar{p}$.
\section{Acknowledgment}
This work is supported by the National Natural Science Foundation of China under
grants Nos. 10805009 and 11035006.
%%%%%%%%%%%%%%%              Reference             %%%%%%%%%%%%%%%%%%%%%

\end{document}